\documentstyle[multicol,aps,prl,epsf]{revtex}
\begin{document}

\draft

\title{
Spin-triplet `f-wave' pairing 
proposed for an organic superconductor (TMTSF)$_2$PF$_6$
}

\author{
Kazuhiko Kuroki, Ryotaro Arita,\cite{JRI}
 and Hideo Aoki 
}

\address{Department of Physics, University of Tokyo, Hongo,
Tokyo 113-0033, Japan}

\date{\today}

\maketitle

\begin{abstract}
By examining how the spin- and/or charge-fluctuation 
exchange can contribute to pairing instabilities, 
we propose that a spin-triplet f-wave-like pairing 
with a $d$-vector perpendicular to the $b$-axis may 
be realized in (TMTSF)$_2$PF$_6$ due to 
(i) a quasi-one-dimensional Fermi surface,
(ii) a coexistence of $2k_F$ charge fluctuations and 
spin fluctuations, and (iii) an anisotropy in spin fluctuations.  
Fluctuation-exchange study for the Hubbard model confirms 
the point (i), while a phenomelogical analysis is given for (ii) and (iii).   
The proposed pairing is consistent with various experiments.
\end{abstract}

\medskip


\begin{multicols}{2}
\narrowtext
Spin-triplet pairing is conceptually fascinating, but 
there seem to be few examples.   
Recently an organic superconductor, (TMTSF)$_2$PF$_6$, 
has attracted much attention since a triplet pairing is suggested 
from an observation of large $H_{c2}$\cite{Lee2} as well as 
from a Knight shift experiment\cite{Lee}, 
while the absence of Hebel-Slichter peak and 
the power-law decay of $T_1^{-1}$ below $T_c$\cite{Lee} 
suggest an anisotropic pairing with nodes in the gap.

If triplet pairing is indeed realized in (TMTSF)$_2$PF$_6$,
its mechanism is a challenging theoretical puzzle: 
in the pressure-temperature phase diagram for this material 
the superconductivity lies right next to the $2k_F$ 
spin density wave (SDW), so that if one seeks an electronic origin, 
a spin-singlet d-wave-like pairing mediated by spin fluctuations 
is most naturally expected as proposed by several authors.
\cite{Shimahara,Kuroki,KK} 
If on the other hand one assumes a phonon-mediated attractive interaction,
triplet pairing, with nodes in the gap in general, 
would seem less favorable compared to singlet s-wave pairing without nodes.  
Recently, Kohmoto and Sato\cite{KS} have proposed that this 
difficulty in the phonon mechanism may be circumvented for TMTSF compounds, 
where they show that the quasi-one dimensionality of the Fermi surface, 
along with the presence of spin fluctuations, 
makes a triplet p-wave pairing without nodes on the Fermi surface\cite{HF}
dominate over s-wave pairing. However, it is not clear if such a nodeless 
gap can be reconciled with the absence of Hebel-Slichter peak and 
the power-law $T_1^{-1}$ in (TMTSF)$_2$PF$_6$.  

If we turn to another prominent 
candidate for triplet superconductivity accompanied by SDW
fluctuations,\cite{Sidis} Sr$_2$RuO$_4$, Takimoto recently proposed that 
{\it charge} fluctuations (or more precisely orbital fluctuations) 
should arise from repulsions between 
degenerate 4d orbitals, and that the coexistence of spin and charge 
fluctuations may lead to a triplet pairing.\cite{Takimoto} 
This makes us recall an experimental fact
that a $2k_F$ charge density wave (CDW) 
actually {\it coexists} with the SDW in (TMTSF)$_2$PF$_6$ 
as suggested from X-ray diffuse scattering.\cite{PR} 
In another theory for Sr$_2$RuO$_4$, Sato and Kohmoto,\cite{SK} and 
independently Kuwabara and Ogata\cite{KO}, have proposed that 
{\it anisotropy} of the spin fluctuations, known to be present
experimentally,\cite{Mukuda} may give rise to a triplet p-wave pairing. 
The anisotropy of spin fluctuations is also present 
in (TMTSF)$_2$PF$_6$.\cite{Mortensen}
Moreover, Sr$_2$RuO$_4$ has two quasi-1D Fermi surfaces (although they are 
weakly hybridized to result in two 2D Fermi surfaces), 
so the ruthenate seems to share several features with 
the TMTSF compound, although the strong charge fluctuation employed in 
Takimoto's mechanism is yet to be detected experimentally in Sr$_2$RuO$_4$.

However, the triplet pairing mechanism of (TMTSF)$_2$PF$_6$ cannot be 
the same with that of Sr$_2$RuO$_4$ since $\vec{d}$ 
(the {\it d}-vector characterizing the triplet pairing) $\perp\vec{z}$ 
(easy axis of the spins) 
is experimentally suggested in the former,\cite{Lee2,Lee} 
while $\vec{d}\parallel\vec{z}$ in the latter.\cite{Ishida}
In the present paper, we propose that a triplet {\it f-wave-like} pairing 
with $\vec{d}\perp\vec{z}$ can take place in (TMTSF)$_2$PF$_6$ 
due to a {\it combination} of (i) the quasi-one-dimensionality of the 
Fermi surface, (ii) coexistence of $2k_F$ spin and charge fluctuations, 
and (iii) the anisotropy in the spin fluctuations.
In the first part of the paper, 
we focus on how the quasi-one-dimensionality works favorably 
for the triplet pairing, and perform a 
fluctuation-exchange (FLEX)\cite{Bickers} calculation 
for the on-site repulsion Hubbard model 
on a lattice for (TMTSF)$_2$PF$_6$. 
Then, in the second part, we discuss phenomelogically how 
the triplet pairing can become competitive against 
the singlet when charge fluctuations coexist with spin fluctuations.  
We finally point out that the anisotropy in the spin fluctuations 
should further favor triplet pairing with $\vec{d}\perp\vec{z}$.

\begin{figure}
\begin{center}
\leavevmode\epsfysize=35mm \epsfbox{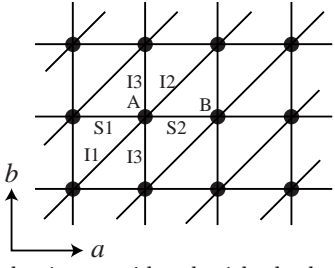}
\caption{The lattice considered with 
the hopping integrals taken to be 
$t_{S1}=-2.8$, $t_{S2}=-2.5$, $t_{I1}=+0.2$, $t_{I2}=+0.5$, 
$t_{I3}=-0.5$ in units of a typical energy scale, $100$ meV, for organics 
after ref.\protect\cite{Ducasse} 
}
\label{fig1}
\end{center}
\end{figure}

We first consider the on-site Hubbard model,
${\cal H}=\sum_{\langle i,j \rangle \sigma} 
t_{ij}c^{\dagger}_{i\sigma}c_{j\sigma}
+U\sum_i n_{i \uparrow}n_{i \downarrow}$, in the hole picture on  
a quasi-1D lattice ($|t_S| > |t_I|$) depicted in Fig.\ref{fig1}.
There are $n=0.5$ holes per site. 
Since sites A and B are inequivalent for $t_{S1}\neq t_{S2}$ and 
$t_{I1}\neq t_{I2}$ (dimerization of the molecules),
we adopt the two-band version of the FLEX\cite{Koikegami,Kontani} 
(although we shall see that the dimerization is not essential in our argument).

For later discussions, 
we first recapitulate the one-band version of FLEX in a general fashion, 
where we proceed: 
(i) Dyson's equation is 
solved to obtain the renormalized
\noindent
Green's function $G(k)$, 
where $k$ is a shorthand for the wave vector ${\bf k}$ and 
the Matsubara frequency, $i\epsilon_n$, 
(ii) the fluctuation-exchange interaction $V^{(1)}(q)$, given as,\cite{Vcom}
\begin{equation}
V^{(1)}(q)=\frac{1}{2}V^{zz}_{\rm sp}(q)+V^{+-}_{\rm sp}(q)+
\frac{1}{2}V_{\rm ch}(q),
\label{v1}
\end{equation}  
consists of the contribution from longitudinal $(zz)$ and 
transverse $(+-)$ spin fluctuations (sp) and that from charge 
fluctuations (ch). 
For the on-site Hubbard model in particular, 
$V^{zz}_{\rm sp}=V^{+-}_{\rm sp}(\equiv V_{\rm sp})=
U^2\chi^{\rm sp}$ and $V_{\rm ch}=U^2\chi^{\rm ch}$, 
where the spin and the charge susceptibilities are given as 
$\chi^{\rm sp}(q)=\chi^{\rm irr}(q)/[1-U\chi^{\rm irr}(q)]$ and 
$\chi^{\rm ch}(q)=\chi^{\rm irr}(q)/[1+U\chi^{\rm irr}(q)]$, 
respectively, using the irreducible susceptibility 
$\chi^{\rm irr}(q)= -\frac{1}{N}\sum_k G(k+q)G(k)$
($N$:number of $k$-point meshes).
(iii) $V^{(1)}$ then brings about the self-energy, 
$\Sigma(k)=\frac{1}{N}\sum_{q} G(k-q)V^{(1)}(q)$,
which is fed back to Dyson's equation, 
and the self-consistent loop is repeated until convergence is attained.

$T_c$ is the temperature where the eigenvalue $\lambda$ of 
the following {\'E}liashberg equation 
for the superconducting order parameter $\phi(k)$ reaches unity.
\begin{eqnarray}
\lambda_\mu\phi_\mu(k)&=&-\frac{T}{N}
\sum_{k'}
\phi_\mu(k')|G(k')|^2 V_\mu^{(2)}(k-k')
\label{eliash}
\end{eqnarray}
Here, the pairing interaction $V_\mu^{(2)}(q)$ is given as  
\begin{eqnarray}
V_s^{(2)}(q)=\frac{1}{2}V^{zz}_{\rm sp}(q)+V^{+-}_{\rm sp}(q)
-\frac{1}{2}V_{\rm ch}(q)
\label{pairs}
\end{eqnarray}
for singlet pairing, 
\begin{eqnarray}
V_{t\perp}^{(2)}(q)=
-\frac{1}{2}V^{zz}_{\rm sp}(q)-\frac{1}{2}V_{\rm ch}(q)
\label{pairt1}
\end{eqnarray}
for triplet pairing with total $S_z=\pm 1$ ($\vec{\it d}\perp\vec{z}$), and 
\begin{eqnarray}
V_{t\parallel}^{(2)}(q)=\frac{1}{2}V^{zz}_{\rm sp}(q)-V^{+-}_{\rm sp}(q)
-\frac{1}{2}V_{\rm ch}(q)
\label{pairt0}
\end{eqnarray}
for triplet pairing with $S_z=0$ ($\vec{\it d}\parallel\vec{z}$).
In the on-site Hubbard model, 
$V_{\rm sp} \gg V_{\rm ch}$ is satisfied,
so that $|V_t^{(2)}|\simeq (1/3)|V_s^{(2)}|$ holds with 
$V_{t\parallel}^{(2)}=V_{t\perp}^{(2)}\equiv V_t^{(2)}$.  

In the two-band version of FLEX,  
$G$, $\chi$, $\Sigma$, and $\phi$ all become $2\times 2$ matrices,
whose elements are denoted as $G_{\alpha\beta}$ etc 
with $\alpha,\beta=$A or B in the site representation, which 
may be converted to 
the band representation with a unitary transform.  
Since the Fermi surface lies in the lower band for quarter filling, 
we concentrate on Green's function and the order parameter 
in that band, denoted as $G$ and $\phi_s, \phi_t$, respectively.
As for the spin susceptibility, we diagonalize the 
$2\times 2$ matrix $\chi^{\rm sp}$ and 
concentrate on the larger eigenvalue, denoted as $\chi$.  
To ensure convergence at low temperatures in the two-band system 
we had to take $64\times 64$ $k$-points and 
$\epsilon_n$ from 
$-(2N_c-1)\pi T$ to $(2N_c-1)\pi T$ with $N_c$ up to 8192.  

\begin{figure}
\begin{center}
\leavevmode\epsfysize=62mm \epsfbox{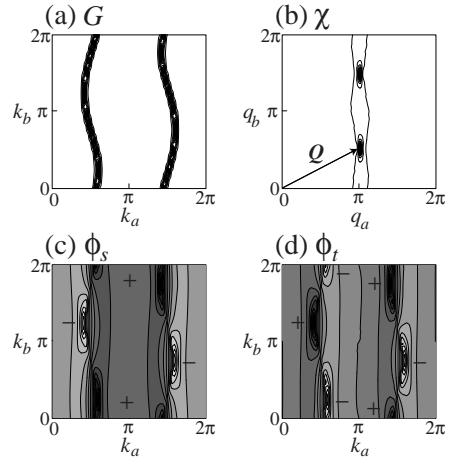}
\caption{Contour plots of 
$|G({\bf k},i\pi k_B T)|^2$ (a), $\chi({\bf q},0)$ (b), 
$\phi_s({\bf k},i\pi k_B T)$ (c), and $\phi_t({\bf k},i\pi k_B T)$ (d)
for $n=0.5$, $U=8$, and $T=0.015$. 
}
\label{fig2}
\end{center}
\end{figure}

In Fig.\ref{fig2}, we show contour plots of  $|G({\bf k},i\pi k_B T)|^2$(a),
$\chi({\bf k},0)$ (b), $\phi_s({\bf k},i\pi k_B T)$ (c), 
and $\phi_t({\bf k},i\pi k_B T)$ (d) for $T=0.015$. 
The Fermi surface as identified from the 
ridge in $|G({\bf k})|^2$ is a pair of warped quasi 1D pieces. 
The spin susceptibility 
$\chi({\bf q},0)$ has a peak at ${\bf Q}\simeq(\pi,\pi/2)$ 
(or $(\pi/2,\pi/2)$ in the unfolded Brillouin zone in the absence of 
dimerization), 
as expected from the nesting vector and in agreement 
with experimental results.\cite{Delrieu,Kawamura}
The singlet pairing order parameter is seen to change sign in such a way that 
(i) $\phi_s({\bf k})=\phi_s(-{\bf k})$, and (ii) 
$\phi_s({\bf k})$ has opposite signs across the nesting vector {\bf Q} so   
that the repulsive $V_s^{(2)}({\bf Q})$ (eq.(\ref{pairs})) 
acts as an attractive interaction in the gap equation.
We call the singlet pairing a `d-wave' in that 
the sign of $\phi_s({\bf k})$ changes like $+-+-$ 
if we rotationally scan the Fermi surface, which is consistent with 
previous studies\cite{Shimahara,Kuroki,KK}.

For the triplet pairing, by contrast, 
$V^{(2)}_t({\bf Q})$ is attractive (eq.(\ref{pairt1}) or (\ref{pairt0})
with $V_{\rm sp}^{+-}=V_{\rm sp}^{zz}$), 
so that the order parameter should have the same sign across ${\bf Q}$.  
This requirement, along with the condititon for a triplet order parameter
$\phi_t({\bf k})=-\phi_t(-{\bf k})$, 
can be satisfied by adding extra nodal lines along 
$k_a\sim 0$ and $k_a\sim \pi$ (mod $2\pi$). 
We call this pairing an `f-wave' in that 
$\phi_s$ behaves this time like $+-+-+-$ along the Fermi surface.

\begin{figure}
\begin{center}
\leavevmode\epsfysize=35mm \epsfbox{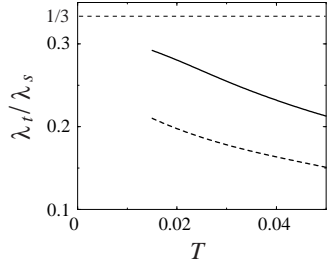}
\caption{$\lambda_t/\lambda_s$  plotted as a function of temperature 
for the parameter values adopted in Fig.\protect\ref{fig2} (solid line),
or for $t_{I2}=-t_{I3}=1.2$ (dashed line).
}
\label{fig3}
\end{center}
\end{figure}

A virtue of the quasi-one dimensionality is that the magnitudes, 
$|\phi_s({\bf k})|$ and $|\phi_t({\bf k})|$, are 
almost identical around the Fermi surface 
as seen from Figs.\ref{fig2}(a) and (c),(d). 
In fact, the `f-wave' here is to `d-wave' what 
p-wave is to s-wave in Kohmoto-Sato's picture\cite{KS}
in that a singlet is converted into a triplet 
by introducing extra nodes that do not affect
$|\phi({\bf k})|$ on the Fermi surface.
In such a situation, the difference between $\lambda_s$ and $\lambda_t$
comes almost entirely 
from the difference between $|V_s^{(2)}({\bf Q})|$ and $|V_t^{(2)}({\bf Q})|$ 
in the {\'E}liashberg equation (\ref{eliash}).

Our result shows that 
the ratio $\lambda_t/\lambda_s$ indeed 
tends to $1/3$ at low temperatures
(Fig.\ref{fig3}; solid line), which reflects the ratio 
$|V_t^{(2)}({\bf Q})/V_s^{(2)}({\bf Q})|\simeq 1/3$ 
in the on-site Hubbard model.  
The ratio approaches to $1/3$ 
as the temperature is lowered, because the ridge in $|G|^2$ 
sharpens so that the triplet pairing, 
with the order parameter vanishing around
$k_a=0, \pi$, becomes more favorable.  
We can also confirm that quasi-1D is exploited in realizing
$\lambda_t/\lambda_s\simeq 1/3$ 
by pushing the system toward 2D with larger value of $t_{I2}$ and $|t_{I3}|$, 
for which
the ratio $\lambda_s/\lambda_t$ deviates from 1/3 even at low temperatures 
(Fig.\ref{fig3}; dashed line).

We have so far seen that the difference in $\lambda$ between 
singlet `d' and triplet `f' can directly reflect the difference between 
$|V_s^{(2)}({\bf Q})|$ and $|V_t^{(2)}({\bf Q})|$ 
in a quasi-1D system.  The `f-wave' proposed here 
is an appealing candidate, because it can 
account for both experimentally suggested triplet pairing 
and the nodes in the superconducting gap.  
However, even for a quasi-1D system, 
`f' is only 1/3 competitive against `d' as far as the on-site 
repulsion Hubbard model is concerned --- 
to make `f' dominate over `d', we need to have 
$|V_t^{(2)}({\bf Q})|>|V_s^{(2)}({\bf Q})|$. 
So at this stage we depart from the Hubbard model 
to argue phenomelogically 
how some factors in the actual (TMTSF)$_2$PF$_6$ 
that are not taken into account in the simple Hubbard model 
can indeed make `f' competitive against `d'. 

An important experimental fact for (TMTSF)$_2$PF$_6$ 
that cannot be explained by the on-site Hubbard model is that 
a $2k_F$ CDW actually coexists with the $2k_F$ SDW\cite{PR}.  
The coexistence of spin and charge fluctuations 
can favor triplet pairing as pointed out 
for Sr$_2$RuO$_4$ by Takimoto mentioned above.\cite{Takimoto}
This can be seen in eqs.(\ref{pairs}),(\ref{pairt1}), 
where an increase in $V_{\rm ch}$ 
enhances $|V_t^{(2)}|$ and suppresses $|V_s^{(2)}|$ 
(as far as $V_{\rm sp}>3V_{\rm ch}$ for isotropic spin fluctuations 
assumed for the time being).  Now, if we take 
the coexistence of $2k_F$ SDW and $2k_F$ CDW in (TMTSF)$_2$PF$_6$
to be $V_{\rm sp}({\bf Q}) \simeq V_{\rm ch}({\bf Q})$,
eqs.(\ref{pairs}) and (\ref{pairt1}) dictate that 
$|V_t^{(2)}({\bf Q})|\simeq |V_s^{(2)}({\bf Q})|$, so 
`f' does indeed compete with `d', but the competition is still subtle. 

Is there any mechanism that further favors the triplet pairing?  
Magnetic anisotropy is, in our view, one.  
It has actually been revealed experimentally for (TMTSF)$_2$PF$_6$ that 
the SDW has an easy axis in the $b$-direction,\cite{Mortensen} 
which implies that 
$V^{zz}_{\rm sp}({\bf Q})>V^{+-}_{\rm sp}({\bf Q})$ is satisfied
for $z$ taken to be $\parallel b$. 
In such a situation, the `f' pairing is more favorable 
in the $S_z=\pm 1$ channel since $|V^{(2)}_{t\perp}|>|V^{(2)}_{t\parallel}|$.
The condition for `f' dominating over `d' now reads 
$|V^{(2)}_{t\perp}({\bf Q})|>V^{(2)}_s({\bf Q})$, or
\begin{equation}
V_{\rm ch}({\bf Q})>V^{+-}_{\rm sp}({\bf Q})
\label{condition}
\end{equation}
from eqs.(\ref{pairs}),(\ref{pairt1}).
This last condition should be satisfied in 
(TMTSF)$_2$PF$_6$ because the spins do not order in the 
transverse direction even in the SDW phase, while the charges do.
We stress that $\vec{d}\perp\vec{z}$ with $\vec{z}\parallel\vec{b}$ 
is consistent\cite{Lebed} with the experimental result: it is when 
the magnetic field is applied {\it parallel to the $b$-axis} that 
(i)$H_{c2}$ becomes largest at low temperatures,\cite{Lee2} 
and (ii) the Knight shift is unchanged across $T_c$\cite{Lee}.

\begin{figure}
\begin{center}
\leavevmode\epsfysize=40mm \epsfbox{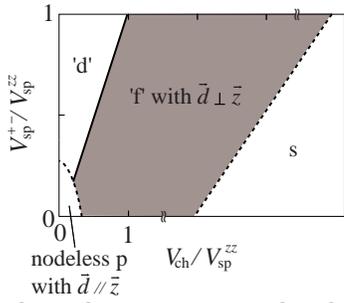}
\caption{A phase diagram against the 
charge/spin axis ($V_{\rm ch}/V^{zz}_{\rm sp}$) and the 
spin anisotropy axis ($V^{+-}_{\rm sp}/V^{zz}_{\rm sp}$).  
The solid line is according to eq.(\ref{condition}), 
while dashed lines are schematic.
}
\label{fig4}
\end{center}
\end{figure}

The mechanism in which the anisotropy of the spin fluctuations 
favors triplet pairing is reminiscent of the one proposed 
in refs.\cite{SK,KO} for Sr$_2$RuO$_4$, 
but a crucial difference is that 
refs.\cite{SK,KO}, which do not consider charge fluctuations, 
conclude a p-wave pairing with $S_z=0$ in agreement with the 
experimental results suggesting $\vec{\it d}\parallel\vec{z}$ in 
Sr$_2$RuO$_4$.\cite{Ishida} 
Let us see how this would occur in the present context.
If $V^{zz}_{\rm sp}>(2V^{+-}_{\rm sp}+V_{\rm ch})$,
we can see from eq.(\ref{pairt0}) that $V^{(2)}_{t\parallel}({\bf Q})$ becomes
{\it repulsive}, which 
will mediate a triplet pairing having an order parameter with
opposite signs across ${\bf Q}$. This requirement, along with 
the triplet condition $\phi_t({\bf k})=-\phi_t(-{\bf k})$, 
can be satisfied by putting nodes only at $k_a\simeq 0$ and $k_a\simeq \pi$,
thereby making the order parameter nodeless on the Fermi surface 
as in refs.\cite{KS,HF}.  Specifically, 
if $V_{\rm ch}/V^{zz}_{\rm sp}$ 
{\it and} $V^{+-}_{\rm sp}/V^{zz}_{\rm sp}$ are both sufficiently small, 
the pairing interactions 
$V^{(2)}_s$ (favoring `d'), $V^{(2)}_{t\perp}$ (`f'),
and $V^{(2)}_{t\parallel}$ (p) will all have similar magnitudes,
so the p-wave pairing, with no nodes on the Fermi surface, 
should dominate over the others. 
Thus, the `f' pairing is not realized unless $V_{\rm ch}/V^{zz}_{\rm sp}$ 
is significant even if eq.(\ref{condition}) is satisfied 
(see Fig.\ref{fig4}).

In this context, a possibly related problem is the pairing symmetry in 
(TMTSF)$_2$ClO$_4$, another candidate for a triplet superconductor.
For this compound an NMR experiment suggests
a presence of nodes on the Fermi surface,\cite{Takigawa} 
while a recent thermal conductivity 
\noindent
measurement suggests a nodeless gap.\cite{Belin} 
If we adopt the latter result, the nodeless p-wave pairing 
should become a strong candidate. Then a comparison of 
the magnitude of the charge fluctuation as well as the direction of 
${\vec d}$ between (TMTSF)$_2$ClO$_4$ 
and (TMTSF)$_2$PF$_6$ will be a crucial test.

Having discussed the lower bound of $V_{\rm ch}/V^{zz}_{\rm sp}$ for 
`f' pairing, how about an upper bound?  
When $V_{\rm ch}/V^{zz}_{\rm sp}\gg 1$,  
{\it singlet s-wave} pairing with 
$\phi_s({\bf k})\sim{\rm constant}$ should enter 
as the dominant pairing.
This is because $V_s^{(2)}$ becomes {\it attractive} for 
$V_{\rm ch} > (V^{zz}_{\rm sp}+2V^{+-}_{\rm sp})$, 
so that $\phi_s({\bf k})$ no longer has to change sign.  
As  $V_s^{(2)}/ V_{t\perp}^{(2)}$ tends to unity with the increase of 
charge fluctuations, the `f', 
with its nodes on the Fermi surface, 
should thus give way to the nodeless s.  
All the above reasoning is schematically summarized 
as a generic phase diagram in Fig.\ref{fig4}. 

An additional bonus from the 
coexistence of strong spin/charge fluctuations and the anisotropic 
spin fluctuations is that 
these effects may serve to enhance the transition temperature 
for triplet pairing.
Namely, a flaw in triplet superconductivity 
mediated by isotropic spin fluctuations is that
the absolute value of the triplet pairing interaction $|V_t^{(2)}|$ 
is only one third of 
the effective interaction $V^{(1)}$ that determines the normal self-energy
as seen from eqs.(\ref{v1}) and (\ref{pairt1}) (or (\ref{pairt0})) with
$V_{\rm sp}^{zz}=V_{\rm sp}^{+-}\gg V_{\rm ch}$. 
This is in contrast with the case of singlet pairing, where 
$V_s^{(2)}$ is nearly identical to $V^{(1)}$. 
Since a large self-energy correction results in a short quasi-particle 
lifetime, $V^{(1)}$ suppresses $T_c$ 
while $V^{(2)}$ enhances it, and 
$V^{(1)}\simeq 3|V_t^{(2)}|$ in the Hubbard model generally results in a 
$T_c$, if any, too low to be detected in FLEX calculations.\cite{AKA,KA}
This difficulty is resolved for large $V_{\rm ch}$ and/or small 
$V^{+-}_{\rm sp}$, for which $|V_{t\perp}^{(2)}|$ approaches $V^{(1)}$.

The microscopic origin of the charge fluctuation
remains to be identified. 
Since there is no orbital degeneracy in (TMTSF)$_2$X, the origin 
cannot be the one proposed by Takimoto for Sr$_2$RuO$_4$.
In fact, the mechanism of $2k_F$ SDW-CDW coexistence in (TMTSF)$_2$PF$_6$ 
has been investigated by several authors. 
Some assume electron-lattice coupling,\cite{Kobayashi1,Mazumdar}
while others envisage a 
purely electronic origin in terms of off-site repulsions up to second
nearest neighbors.\cite{Kobayashi2} It would be an interesting future problem 
to investigate microscopically the singlet-triplet competition and 
to evaluate $T_c$ by taking these effects into account.
 
We thank Tetsuya Takimoto for illuminating discussions and 
sending ref.\cite{Takimoto} prior to publication. 
We are also indebted to Mahito Kohmoto, Masao Ogata, 
and Masatoshi Sato for valuable discussions. 
One of us (H.A.) also thanks John Singleton for discussions on the 
SDW/CDW problem.
Numerical calculations were performed at the Computer Center,
University of Tokyo, and at the Supercomputer Center,
ISSP, University of Tokyo.

\end{multicols}

\begin{references}
\bibitem[*]{JRI} Present address: The Japan Research Institute,
Ichibancho, Chiyoda-ku, Tokyo, 102-0082, Japan
\bibitem{Lee2} I.J. Lee {\it et al.}, Phys. Rev. Lett. {\bf 78}, 3555 (1997). 
\bibitem{Lee} I.J. Lee {\it et al.}, preprint (cond-mat/0001332).
\bibitem{Shimahara} H. Shimahara, J. Phys. Soc. Jpn. {\bf 58}, 1735 (1989).
\bibitem{Kuroki} K. Kuroki and H. Aoki, Phys. Rev. B {\bf 60}, 3060 (1999).
\bibitem{KK} H. Kino and H. Kontani, J. Low Temp. Phys. {\bf 117}, 317 (1999).
\bibitem{KS} M. Kohmoto and M. Sato, preprint (cond-mat/0001331).
\bibitem{HF} This type of pairing was first considered in Y. Hasegawa and 
H. Fukuyama, J. Phys. Soc. Jpn. {\bf 55}, 3978 (1986).
\bibitem{Sidis} Y. Sidis {\it et al.}, Phys. Rev. Lett. {\bf 83}, 3320 (1999).
\bibitem{Takimoto} T. Takimoto, preprint.
\bibitem{PR} J.P. Pouget and S. Ravy, J. Phys. I (France) {\bf 6}, 1501 (1996).
\bibitem{SK} M. Sato and M. Kohmoto, preprint (cond-mat/0003046).
\bibitem{KO} T. Kuwabara and M. Ogata, preprint (cond-mat/ 0003296).
\bibitem{Mukuda} H. Mukuda {\it at al.}, 
J. Phys. Soc. Jpn. {\bf 67}, 3945 (1998).
\bibitem{Mortensen} K. Mortensen {\it et al.}, Phys. Rev. Lett. {\bf 46}, 
1234 (1981).
\bibitem{Ishida} K. Ishida {\it et al.}, Nature {\bf 396}, 658 (1998);
\bibitem{Bickers} N.E. Bickers, D.J. Scalapino, and S.R. White, 
Phys. Rev. Lett. {\bf 62}, 961 (1989); G.Esirgen and N.E.Bickers, Phys. Rev. 
B {\bf 55}, 2122 (1997).
\bibitem{Ducasse} L. Ducasse {\it et al.}, J. Phys. C {\bf 19}, 3805 (1986).
\bibitem{Koikegami} S. Koikegami, S. Fujimoto, and K. Yamada, J. Phys. Soc. 
Jpn. {\bf 66}, 1438 (1997).
\bibitem{Kontani} H. Kontani and K. Ueda, Phys. Rev. Lett. {\bf 80}, 
5619 (1998).
\bibitem{Vcom} In eqs.(\protect\ref{v1}),(\protect\ref{pairs})
$-$(\protect\ref{pairt0}), we omit the 1st and 2nd order terms,
which are negligible when spin and/or charge fluctuations are strong, 
although we have taken those terms into account in the actual calcuation.
\bibitem{Delrieu} J.M. Delrieu {\it et al.}, J. Phys. (France) {\bf 47}, 839
(1986).
\bibitem{Kawamura} H. Kawamura {\it et al.}, J. Phys. Soc. Jpn. {\bf 55}, 
1364 (1986).
\bibitem{Lebed} A.G. Lebed, K. Machida, and M .Ozaki, preprint 
(cond-mat/0005039).
\bibitem{Takigawa} M. Takigawa, 
H. Yasuoka, and G. Saito, J. Phys. Soc. Jpn. {\bf 56}, 873 (1987)
\bibitem{Belin} 
S. Belin and K. Behnia, Phys. Rev. Lett. {\bf 79}, 2125 (1997)
\bibitem{AKA} R. Arita, K. Kuroki, and H. Aoki, Phys. Rev. B {\bf 60}, 14585 
(1999); J. Phys. Soc. Jpn. {\bf 69}, 1181 (2000).
\bibitem{KA} K. Kuroki and R. Arita, preprint (cond-mat/0004381), 
have recently proposed that a triplet pairing can dominate 
with possibly a finite $T_c$ 
even when mediated by isotropic spin-fluctuations 
in the on-site Hubbard model, 
provided the Fermi surface consists of disconnected pieces 
(as in triangular or honeycomb lattices) 
with the nesting vector lying within each piece. 
\bibitem{Kobayashi1} N. Kobayashi and M. Ogata, J. Phys. Soc. Jpn. {\bf 66}, 
3356 (1997).
\bibitem{Mazumdar} S. Mazumdar {\it et al.}, Phys. Rev. Lett. {\bf 82}, 1522
(1999).
\bibitem{Kobayashi2} N. Kobayashi, M. Ogata, and K. Yonemitsu,
J. Phys. Soc. Jpn. {\bf 67}, 1098 (1998). 
\end{references}
\end{document}